\documentclass{aastex}
\usepackage{spr-astr-addons}
\usepackage{url}

\RequirePackage{color}

\newcommand{\emaila}{JohnZG.Ma@asc-csa.gc.ca}

\begin{document}

\title{Nonlinear ion-acoustic (IA) waves driven in a cylindrically symmetric flow}
\shorttitle{Nonlinear IA waves in a cylindrically symmetric flow}
\shortauthors{Ma}

\author{John Z. G. Ma}
\affil{Space Science Branch, Canadian Space Agency, Canada}
\email{\emaila}

\begin{abstract}
By employing a self-similar, two-fluid MHD model in a cylindrical
geometry, we study the features of nonlinear ion-acoustic (IA) waves
which propagate in the direction of external magnetic field lines in
space plasmas. Numerical calculations not only expose the well-known
three shapes of nonlinear structures (sinusoidal, sawtooth, and
spiky or bipolar) which are observed by numerous satellites and
simulated by models in a Cartesian geometry, but also illustrate new
results, such as, two reversely propagating nonlinear waves, density
dips and humps, diverging and converging electric shocks, etc. A
case study on Cluster satellite data is also introduced.
\end{abstract}


\section{Introduction}

Nonlinear plasma theory and approaches have been developed for more
than half a century \citep{sag69,dav72}. Important problems, such as
the excitation, propagation, and effects of nonlinear waves, have
been extensively studied since 1970s, see, e.g., \citet{inf00}. One
branch lies in cylindrically symmetric plasma systems which are
ubiquitously observed in geo-space by a multitude of measurements
from, e.g., ground-based imagers \citep{pim01}, rockets
\citep{ear89,moo96}, and satellites \citep{pic04,vai04,dek05}.

In the cylindrical frame where $\mathbf{B}$ is along axial
$z$-direction, small-amplitude ion acoustic (IA) waves were firstly
studied \citep{max74,max76}. Then, nonlinear surface waves were
found to shift the radius of a plasma cylinder \citep{gra83,gra84}.
After that, the deformed amplitudes and propagation velocities were
confirmed to be different from those in the semi-infinite limit of
plasmas \citep{gra85,gra86}. In addition, the equations governing
evolution of surface and body waves were obtained \citep{mol87}.
Furthermore, in unmagnetized systems, a set of nonlinear equations
was solved to describe the temporal change of the electron density
in strongly nonlinear surface waves \citep{ste90,yu91}. The work was
followed by a generalized study including the rotation effects of a
time-dependent rigid plasma body \citep{ste92,ste95}. By contrast,
in magnetized systems, nonlinear IA parallel-propagating electric
field structures were recently calculated \citep{shi01,shi05}, and
bipolar electric field structures were verified to be able to
originate from either IA or ion cyclotron (IC) waves \citep{shi08}.

By examining observations by high-resolution satellites (e.g., Wind,
FAST, Polar, Cluster), we found that, in addition to the bipolar
shapes of nonlinear structures, there exist two other well-known
nonlinear electric field envelops: sinusoidal and sawtooth. These
three shapes were first recorded by S3-3 in the 1970s \citep{tem79}.
Since then, they been detected by numerous satellites such as S3-3
\citep{tem82}, Viking \citep{bos88}, Geotail \citep{mat94}, Wind
\citep{bal98}, FAST \citep{erg98,mcf99,mcf03}, Polar
\citep{moz97,fra98,bou99,cat99,fra00}, and Cluster
\citep{pic04,pic05}.

The formation of these three nonlinear structures have been studied
extensively in Cartesian coordinates. For example, \citet{tem79}
firstly reproduced these shapes by solving a set of fluid equations
in the ion-cyclotron (IC) / ion-acoustic (IA) regime. In a unified
work, \citet{lee81} obtained these two important nonlinear
electrostatic waves (it is worth to mention here that the authors
also mentioned a third type of the nonlinear waves: ``ion-acoustic
solitons". By redoing the calculations, we can easily see that it is
not a new type but the simple waves with a longer period for peaks
to occur). Their studies were followed by, e.g.,
\citet{nak96,cha97,das00,jov00,mam02}. Particularly,
\citet{red02,bha02,ma09} not only verified that the nonlinear
structures originate from a coupling between the IC and IA modes,
but also obtained the three waveforms reported firstly by
\citet{tem79}. The authors found that the nonlinear shapes change
from a sinusoidal IC/IA structure of small amplitudes at low Mach
numbers to sawtooth and then bipolar/spiky waveforms of large
amplitudes at higher Mach numbers in parallel propagations.

In cylindrical coordinates, however, there was, to our knowledge,
only one study on the contribution of cylindrically symmetric
plasmas to the emergence and propagation of nonlinear structures
measured by satellites. This study was done by \citet{tri05,tri06}.
The authors offered a matched picture between the Cluster data of
nonlinear waves and the kinetic modeling of the wave excitation in
the perpendicular plane of the local magnetic field $\mathbf{B}$.
Inspired by their work, we pay attention to parallel-propagating
nonlinear waves, aiming at clarifying if or not the well-known three
envelops of nonlinear structures can be driven in cylindrically
symmetric flows, and if possible, obtaining new features of
nonlinear waves in such a system, such as density holes and humps.
The organization of the article is as follows. Section 2 set up a
two-fluid model by employing a set of self-similar MHD equations.
Section 3 introduces a parameterized analysis of nonlinear waves by
numerical calculations. Several features of them are obtained.
Section 4 compares the results with that in a Cartesian frame, and
have a case study on the Cluster data. Section 5 presents a
conclusion.
\begin{figure*}[htb]
\begin{center}
\includegraphics[scale=0.8,angle=0]{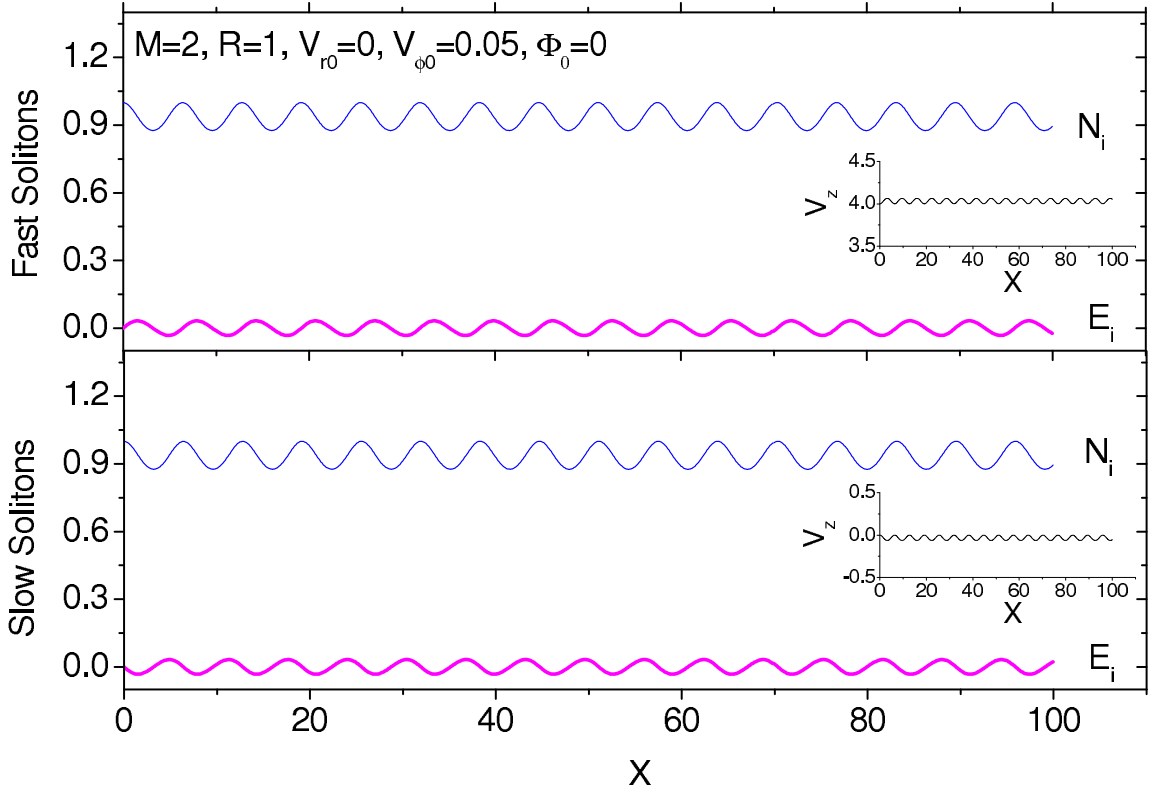}
\end{center}
\caption{(Color online) Ion density $N_{i}$ and wave-field strength
$E_{i}$ of sinusoidal nonlinear waves excited under $V_{\phi0}$=0.05
at $M$=2 and $R$=1. Upper panel: Fast nonlinear wave; Lower panel:
Slow nonlinear wave. Inserted in each panel is the speed of the
simple wave propagating along $B$.}\label{axial-1}
\end{figure*}

\section{Two-fluid model}

In order to provide the most basic picture for the propagation of
electrostatic nonlinear waves driven in cylindrically symmetric
plasma regions, and be able to illustrate clearly the modulation of
nonlinear structures by various input parameters and boundary
conditions, we focus on FAST/Cluster orbits where the plasma $\beta$
is much less than 1 (such as, the auroral acceleration regions, bow
shock, magnetopause), and employ a two-fluid model in a cylindrical
frame $(r,\phi,z)$ with $\mathbf{B}$=$B\hat{\mathbf{e}}_{z}$ (where
$B$ is constant, $\hat{\mathbf{e}}_{z}$ is the unit vector along $z$
axis).

The model takes into account isothermal electron and ion fluids,
with $v_{Te}$$\gg$$v_{Ti}$, where
$v_{Te}$=$\sqrt{2k_{B}T_{e}/m_{e}}$ and
$v_{Ti}$=$\sqrt{2k_{B}T_{i}/m_{i}}$ are the electron and ion thermal
speeds, respectively, in which $k_{B}$ is Boltzmann's constant,
$T_{e}$ and $T_{i}$ are their temperatures (assumed constant),
respectively, and $m_{e}$ and $m_{i}$ are their masses,
respectively. In this study, we neglect the electron inertia because
of $m_{e}$$\ll$ $m_{i}$. Both the momentum equation and the
isothermal equation of state of the electron fluid provide
$N_{e}$=$e^{\Phi}$, where $N_{e}$=$n_{e}/n_{0}$ is the dimensionless
electron density in which $n_{e}$ and $n_{0}$ are the dimensional
electron density and the uniform background plasma density,
respectively, and $\Phi$=$e\varphi/(k_{B}T_{e})$ where $\varphi$ is
the perturbed electrostatic potential.

The ion fluid is described by a set of MHD equations expressing ion
continuity, momentum (containing the equation of state), and the
quasi-neutrality:
\begin{equation}
\left\{\begin{array}{lll} \frac{\partial N_{i}}{\partial
\tau}+\frac{\partial (N_{i}V_{r})}{\partial R}+\frac{\partial
(N_{i}V_{z})}{\partial Z}=-\frac{N_{i}V_{r}}{R}\\
\frac{\partial V_{r}}{\partial \tau}+V_{r}\frac{\partial
V_{r}}{\partial R}+V_{z}\frac{\partial V_{r}}{\partial
Z}=\frac{V_{\phi}^{2}}{R}+V_{\phi}-\zeta\frac{\partial\mathrm{ln}N_{i}}{\partial
R}\\
\frac{\partial V_{\phi}}{\partial \tau}+V_{r}\frac{\partial
V_{\phi}}{\partial R}+V_{z}\frac{\partial V_{\phi}}{\partial
Z}=-\frac{V_{r}V_{\phi}}{R}-V_{r}\\
\frac{\partial V_{z}}{\partial \tau}+V_{r}\frac{\partial
V_{z}}{\partial R}+V_{z}\frac{\partial V_{z}}{\partial
Z}=-\zeta\frac{\partial\mathrm{ln}N_{i}}{\partial Z}\\
N_{i}\approx N_{e}=e^{\Phi}
\end{array}\right.
\label{set-new2}
\end{equation}
in which $N_{i}$=$n_{i}/n_{0}$ (where $n_{i}$ is the ion density),
$\tau$=$\Omega_{i}t$ (where $\Omega_{i}$=$eB/m_{i}$ is the ion
gyro-frequency, and $t$ is time),
$V_{r}$=$u_{r}/c_{s}$,$V_{\phi}$=$u_{\phi}/c_{s}$,$V_{z}$=$u_{z}/c_{s}$
(where $\{u_{r},u_{\phi},u_{z}\}$ is the ion velocity components,
and $c_{s}=\sqrt{k_{B}T_{e}/m_{i}}$ is the ion acoustic speed),
$R$=$r/\rho_{i}$, $Z$=$z/\rho_{i}$ (where
$\rho_{i}$=$c_{s}/\Omega_{i}$ is the ion gyro-radius), and
$\zeta=1+v_{Ti}^{2}/(2c_{s}^{2})$. Note that (1) due to the
symmetric nature of the cylindrical system, all derivative terms
along the $\phi$-direction do not occur; and (2) the parameter $R$
represents the radius of curvature of the flow streamline
intersecting the magnetic field lines on which the equation is going
to be solved.

In this system, linear IA and IC waves can be excited. By
linearizing Eq.(\ref{set-new2}), we obtain $\omega_{1}^{2}$=$\zeta
k^{2}c_{s}^{2}$ (where $k$ is the amplitude of the wave vector
$\mathbf{k}$) in the parallel direction and
$\omega_{2}^{2}$=$\Omega_{i}^{2}$ in the transverse plane.
Superimposing upon these background oscillations, there exist
nonlinear waves the features of which can be obtained by solving a
set of self-similar equations of Eq.(\ref{set-new2}) via introducing
a self-similar parameter $X$ \citep{lee81,shi01}:
\begin{equation}
X=\frac{\alpha_{1}}{M}R+\frac{\alpha_{2}}{M}Z-\tau
\end{equation}
where $M$ is the Mach number, $\alpha_{1}$=$\mathrm{sin}\theta$, and
$\alpha_{2}$=$\mathrm{cos}\theta$ in which $\theta$ is the
inclination angle between the propagation direction and the magnetic
field. Using self-similar transformations, i.e.,
\begin{equation}
\frac{\partial}{\partial\tau}=-\frac{\mathrm{d}}{\mathrm{d}X},\ \
\frac{\partial}{\partial
R}=\frac{\alpha_{1}}{M}\frac{\mathrm{d}}{\mathrm{d}X},\ \
\frac{\partial}{\partial
Z}=\frac{\alpha_{2}}{M}\frac{\mathrm{d}}{\mathrm{d}X}
\end{equation}
We concentrates on parallel-propagating nonlinear waves, i.e.,
$\theta$=0. This means $\alpha_{1}$=0, and $\alpha_{2}$=1.
Therefore, we have
\begin{equation}
\frac{\partial}{\partial R}=0,\ \ \frac{\partial}{\partial
Z}=\frac{1}{M}\frac{\mathrm{d}}{\mathrm{d}X} \label{2partial}
\end{equation}
the first expression of which indicates that $R$ is independent of
$X$. We thus obtain a set of four self-similar equations of
nonlinear waves propagating along $\mathbf{B}$:
\begin{equation}\left\{\begin{array}{lll}
\frac{\mathrm{d}V_{r}}{\mathrm{d}X}-\frac{V_{z}}{M}\frac{\mathrm{d}
V_{r}}{\mathrm{d}X}=-\frac{V_{\phi}^{2}}{R}-V_{\phi}\\
\frac{\mathrm{d}V_{\phi}}{\mathrm{d}X}-\frac{V_{z}}{M}\frac{\mathrm{d}
V_{\phi}}{\mathrm{d}X}=\frac{V_{r}V_{\phi}}{R}+V_{r}\\
\frac{\mathrm{d}V_{z}}{\mathrm{d}X}-\frac{V_{z}}{M}\frac{\mathrm{d}V_{z}}{\mathrm{d}X}=\frac{\zeta}{M}\frac{\mathrm{d}\Phi}{\mathrm{d}X}\\
\left(1-\frac{V_{z}}{M}\right)\frac{\mathrm{d}\Phi}{\mathrm{d}X}-\frac{1}{M}\frac{\mathrm{d}
V_{z}}{\mathrm{d}X}=\frac{V_{r}}{R}
\end{array}\right.
\label{set-new3}\end{equation} in which $R$ behaves only as an input
parameter to represent the existence of geometrical effects (namely,
the centrifugal and Coriolis forces) in curvilinear flows. Note that
these effects are absent in rectilinear systems.

Using boundary conditions $\Phi|_{X=0}=0$, the third equation of
Eq.(\ref{set-new3}) can be integrated to give
\begin{equation}
V_{z}=M\left(1\pm\sqrt{1-\Psi}\right)
\end{equation}
where $\Psi$=2$\zeta\Phi/M^{2}$ and $V_{z}|_{X=0}$ have two initial
values: 0 and 2$M$. This leads to a new set of equations as follows:
\begin{equation}
\left\{\begin{array}{lll} \pm\sqrt{1-\Psi}\frac{\mathrm{d}
V_{r}}{\mathrm{d}X}=\left(\frac{V_{\phi}}{R}+1\right)V_{\phi}\\
\mp\sqrt{1-\Psi}\frac{\mathrm{d}
V_{\phi}}{\mathrm{d}X}=\left(\frac{V_{\phi}}{R}+1\right)V_{r}\\
\pm\left(-\frac{M^{2}}{\zeta}\sqrt{1-\Psi}+\frac{1}{\sqrt{1-\Psi}}\right)\frac{\mathrm{d}\Psi}{\mathrm{d}X}=2\frac{V_{r}}{R}\\
E_{i}=-\frac{1}{M}\frac{\mathrm{d}\Phi}{\mathrm{d}X}=-\frac{M}{2\zeta}\frac{\mathrm{d}\Psi}{\mathrm{d}X}
\end{array}\right.
\label{set-new003}
\end{equation}
where $E_{i}$ is the dimensionless electric field amplitude of the
simple wave with a unit of $E_{0}$=$c_{s}B$.
\begin{figure*}[htb]
\begin{center}
\includegraphics[scale=0.8,angle=0]{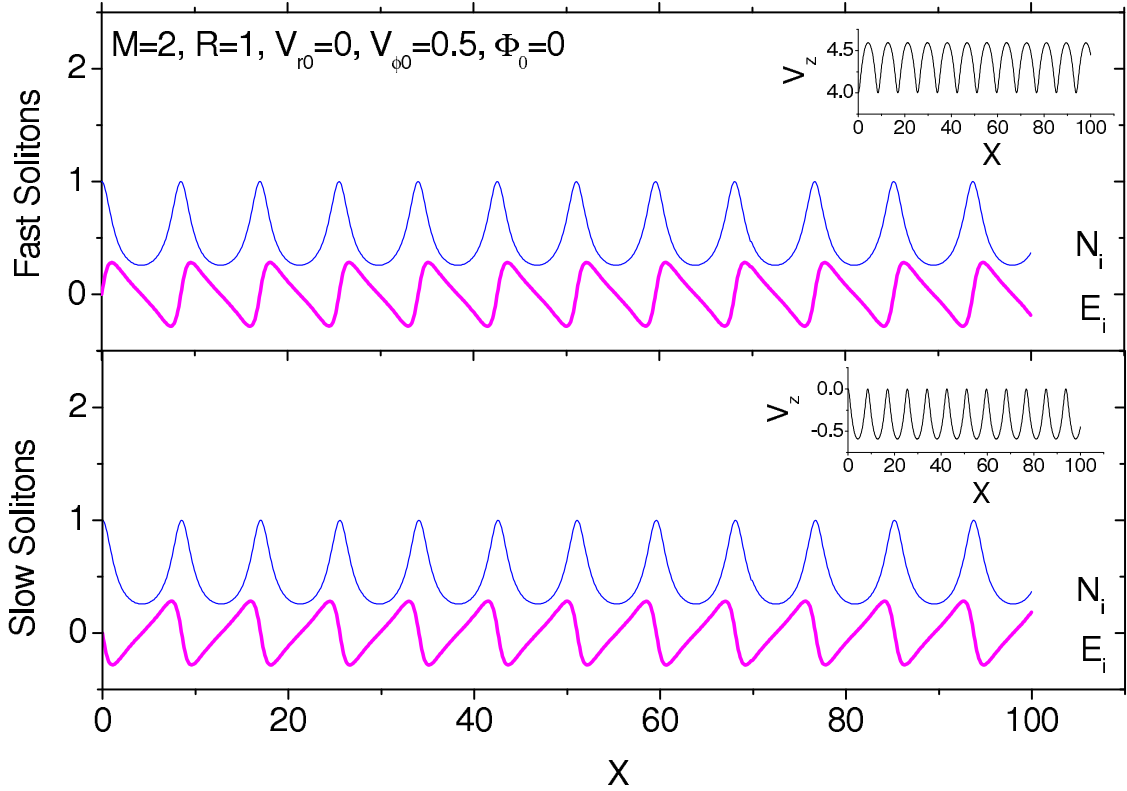}
\end{center}
\caption{(Color online) The same as Fig.\ref{axial-1} but sawtooth
nonlinear waves excited under $V_{\phi0}$=0.5.}\label{axial-2}
\end{figure*}

\section{Parameterized analysis}

\subsection{Prerequisite of the excitation of nonlinear waves}

Cylindrically-symmetric flows are azimuthal with $V_{\phi0}\neq0$ at
$X$=0. Other parameters satisfy equilibrium conditions, i.e.,
$V_{r0}$=$\Phi_{0}$=0. A nonzero $V_{\phi0}$ is important. This can
be seen from the first two equations of Eq.(\ref{set-new003}). They
provide
\begin{equation}
V_{r}\mathrm{d}V_{r}+V_{\phi}\mathrm{d} V_{\phi}=0
\end{equation}
or,
\begin{equation}
V_{r}^{2}+V_{\phi}^{2}=V_{r}^{2}|_{X=0}+V_{\phi}^{2}|_{X=0}=V_{\phi0}^{2}
\end{equation}
Clearly, if $V_{\phi0}$=0, both $V_{r}$ and $V_{\phi}$ are zeros at
any $X$, and thus, the last equation of Eq.(\ref{set-new003}) gives
$\Psi$=0 at any $X$. In this case, no nonlinear waves can develop.

As a result, only under nonzero $V_{\phi0}$ conditions is it
possible to trigger nonlinear processes. Luckily, this condition is
always met naturally in toroidal flows in geospace.
\begin{figure*}[htb]
\begin{center}
\includegraphics[scale=0.8,angle=0]{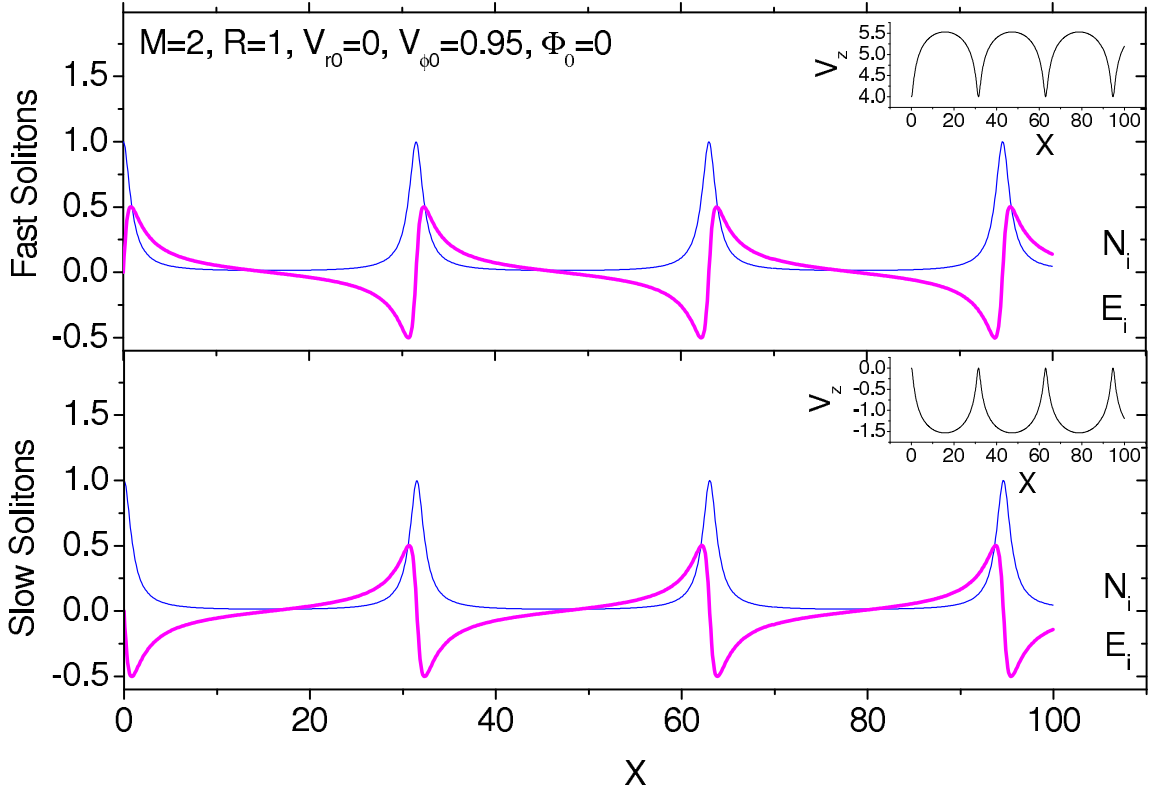}
\end{center}
\caption{(Color online) The same as Fig.\ref{axial-1} but spiky (or
bipolar) nonlinear waves excited under $V_{\phi0}$=0.95. Nonlinear
waves have density holes.}\label{axial-3}
\end{figure*}

\subsection{Evolution of nonlinear waves}

At three different levels of $V_{\phi0}$=0.05 (weak), 0.5 (medium),
0.95 (strong), respectively, Figs.\ref{axial-1}-\ref{axial-3}
illustrate, respectively, the three well-known structures of the
nonlinear wave-field $E_{i}$, namely, sinusoidal, sawtooth and spiky
or bipolar, as well as the accompanied nonlinear-wave density
$N_{i}$, under $M$=2 and $R$=1. Every figure contains two panels,
each of which is inserted into a $V_{z}$-curve to express the
propagating speed of a nonlinear wave along the axial direction.
Note that the fast wave is always propagating in the direction of
$\mathbf{B}$ ($V_{z}$$\geq$0), while the other in the opposite
direction ($V_{z}$$\leq$0).

The three figures illustrate that the weak flow drives sinusoidal
nonlinear structures with small amplitudes ($|E_{i}|_{max}$ =0.03),
the medium flow drives sawtooth structures with medium amplitudes
($|E_{i}|_{max}$=0.3), and the strong flow drives spiky (bipolar)
nonlinear waves with high amplitudes ($|E_{i}|_{max}$=0.5). This
result reproduces the results obtained by \citet{red02,bha02} in
Cartesian coordinates. This indicates that the features of nonlinear
waves are strengthened increasingly with stronger drifts of
azimuthal flows. Especially, in the last case, the two panels in
Fig.\ref{axial-3} reveal diverging shocks$^{18}$ (a negative
electric field followed by a positive one) of the fast
nonlinear-wave packet in the upper panel, and converging
shocks$^{43}$ (a positive electric field followed by a negative one)
of the slow packet in the lower panel. In addition, more
calculations with a changing $M$ expose that under $M$$>$1, the
nonlinear-wave density $N_{i}$ is never larger than 1, but goes to a
minimum which is smaller at a faster drift, meaning density holes
are formed and their boundaries result in nonlinear waves. For
example, the hole is 0.876 in amplitude with a smaller
$|E_{i}|_{max}$ at $V_{\phi0}$=0.05, while it becomes 0.015 with a
larger $|E_{i}|_{max}$ at $V_{\phi0}$=0.95. We note that $V_{\phi0}$
cannot be larger than 1, that is, the azimuthal flow speed is unable
to surpass the local acoustic speed.

\subsection{Influence of $R$.}

\begin{figure*}[htb]
\begin{center}
\includegraphics[scale=0.8,angle=0]{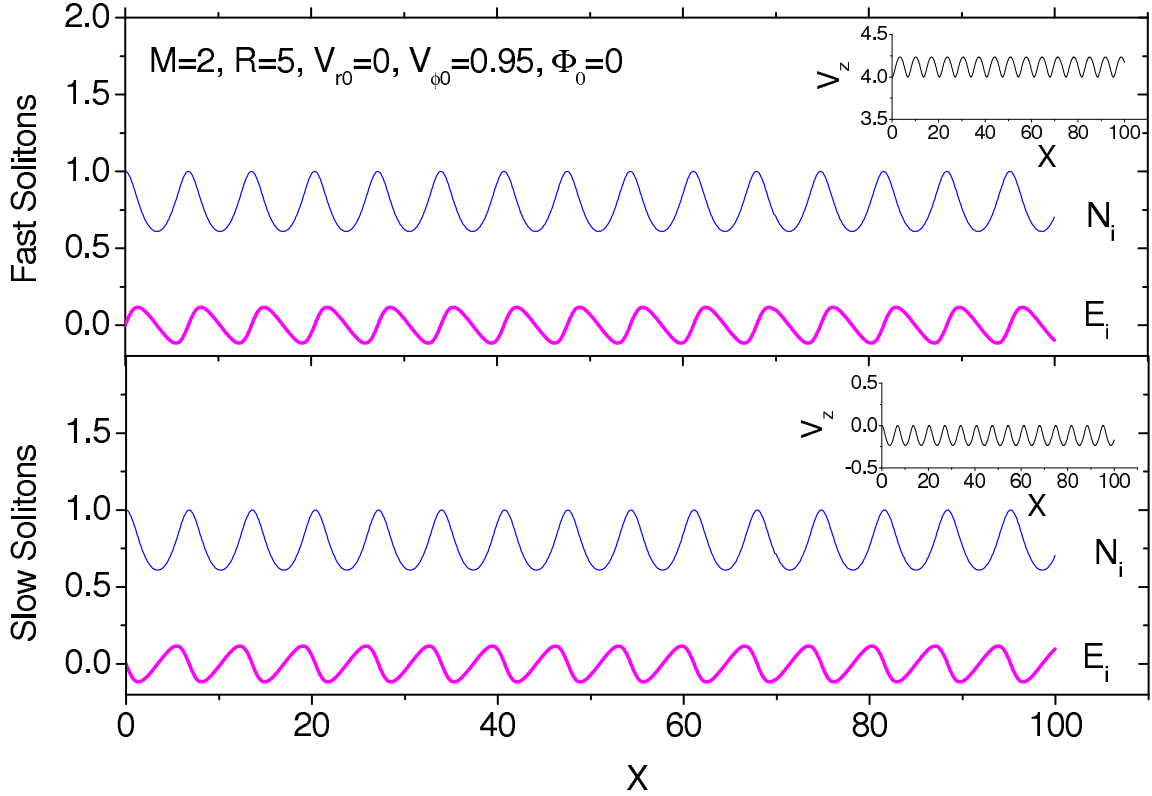}
\end{center}
\caption{(Color online) The same as Fig.\ref{axial-3} but sawtooth
nonlinear waves excited at a distant radial position
$R$=5.}\label{axial-4}
\end{figure*}

In a cylindrical system, the frame effects produced by the
centrifugal and Coriolis terms in the momentum equations decrease
with radius. They should have a direct influence on the structure of
nonlinear waves excited in the azimuthal flows. Fig.\ref{axial-4}
expose the role played by $R$. At a larger radius $R$=5 compared to
Fig.\ref{axial-3}, the nonlinear waves change their appearances from
a spiky (bipolar) shape to a sawtooth one, with a lower amplitude
$|E_{i}|_{max}$=0.12 but a higher frequency. At another radius,
$R$=10, a calculation shows that sinusoidal shape emerges, with a
very small amplitude of $|E_{i}|_{max}$=0.06. This indicates that
the features of nonlinear waves are weakened when going farther from
the center of the cylindrically symmetric flow. It is thus
reasonable to propose that it is easier to detect nonlinear waves
closer to the symmetric axis where the influence of the centrifugal
and Coriolis forces are more dominant.

\subsection{Criterion for density holes and humps.}

\begin{figure*}[htb]
\begin{center}
\includegraphics[scale=0.8,angle=0]{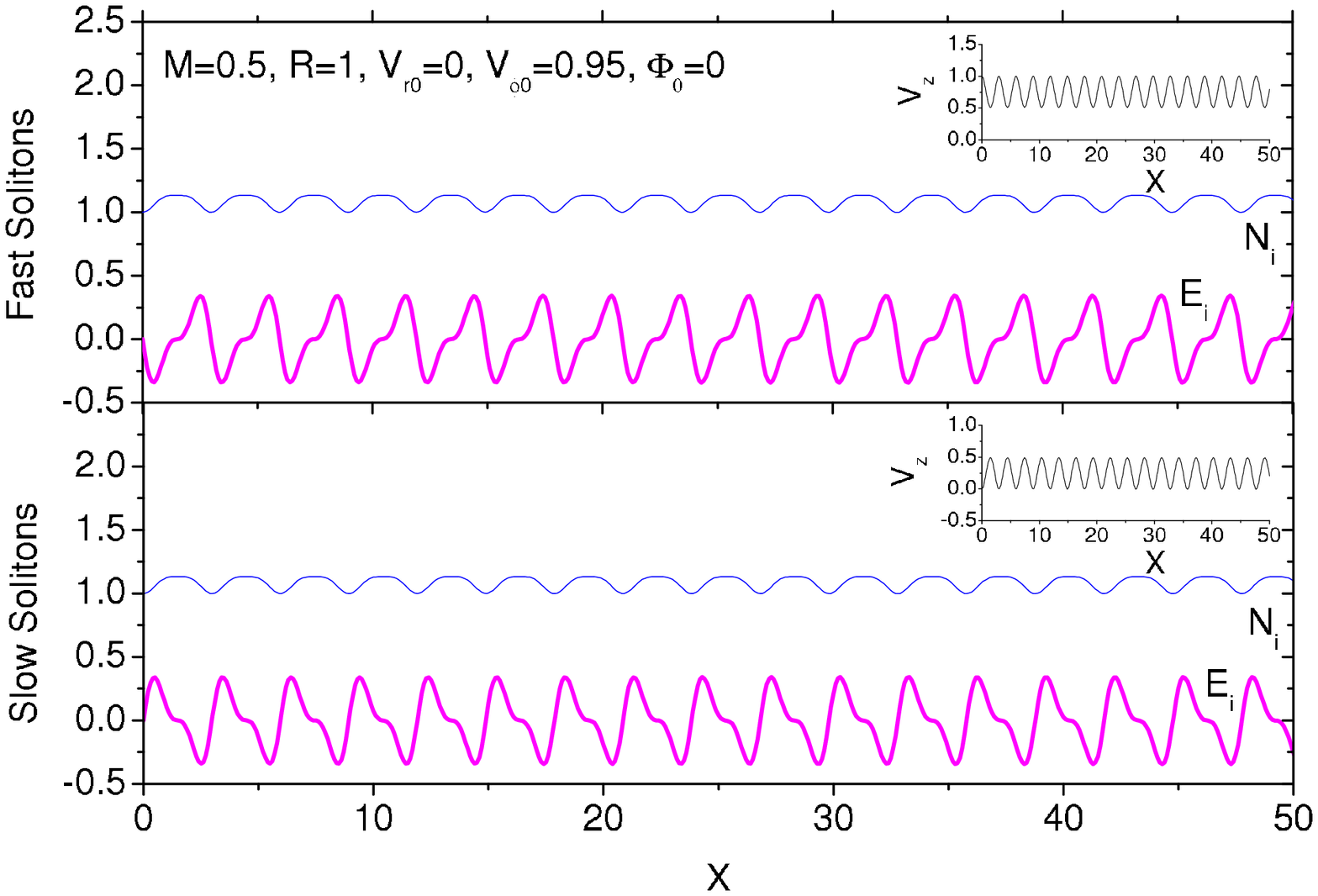}
\end{center}
\caption{(Color online) The same as Fig.\ref{axial-3} but
higher-frequency spiky (or bipolar) nonlinear waves excited by a low
Mach-number $M$=0.5. Nonlinear waves have density
humps.}\label{axial-5}
\end{figure*}

The Mach number cannot always be larger than 1,$^{36}$ referring to
that $M$$<$1 is also possible. Fig.\ref{axial-5} illustrates a case
under conditions of Fig.\ref{axial-3} but at $M$=0.5. Interestingly,
different from the density holes, there are now density humps coming
into being: the nonlinear-wave density $N_{i}$ is never smaller than
1, and goes to a maxmum 1.133. In addition, the bipolar nonlinear
wave becomes denser with a much higher frequency of 3, than that of
0.04 in Fig.\ref{axial-3}. However, the amplitude of $E_{i}$
decreases from 0.5 to 0.35, meaning the nonlinear feature is
weakened. Other calculations confirm that density holes occur at
$M$$>$1, while density humps appear at $M$$<$1.

\section{A case study}

\begin{figure}[htb]
\begin{center}
\includegraphics[scale=0.4,angle=0]{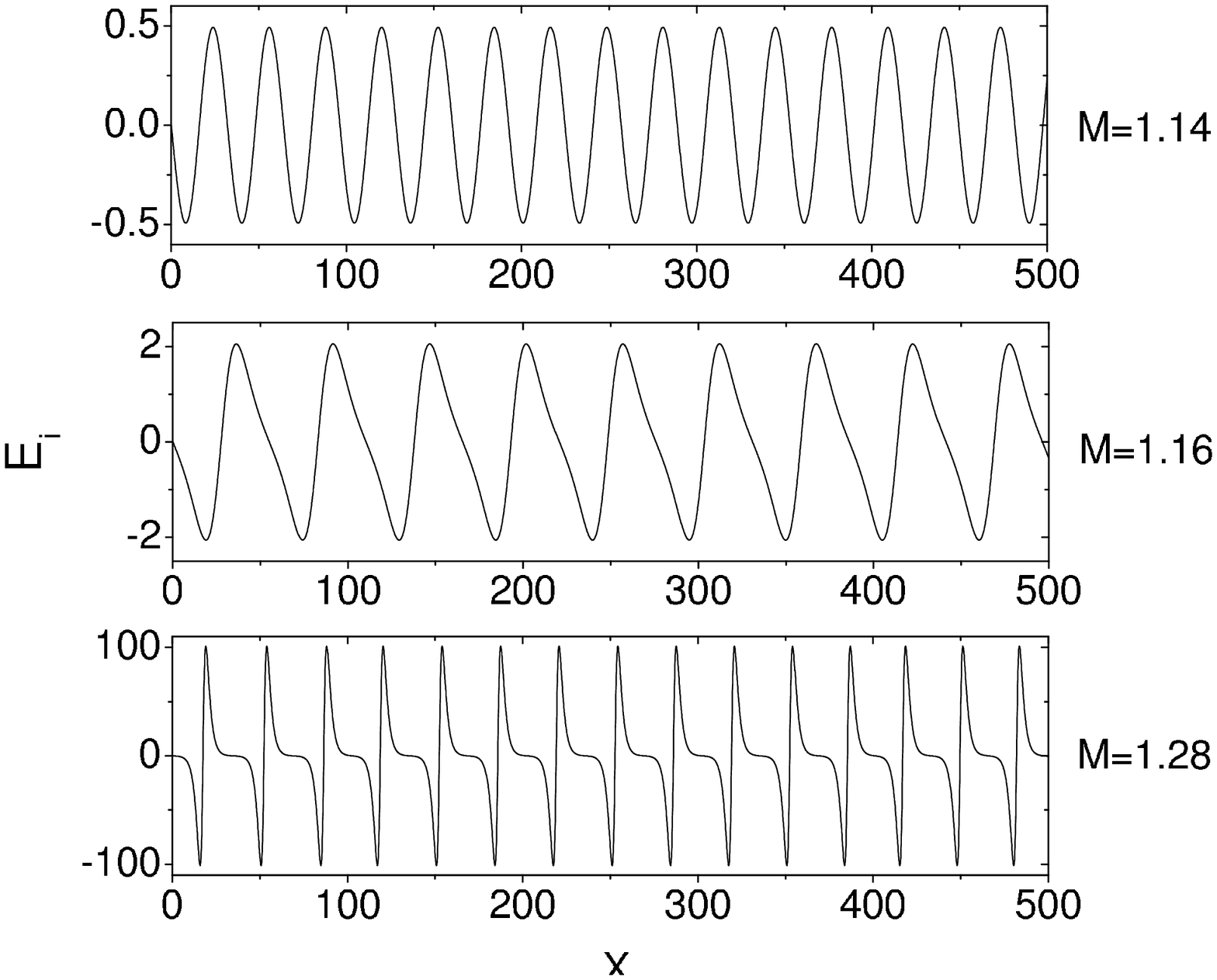}
\end{center}
\caption{Parallel-propagating nonlinear waves developed from a
rectangular flow. Although there still exist three types of shapes
(sinusoidal, sawtooth, and spiky or bipolar), only diverging shocks
can be evolved from such a system modulated by the Mach number
$M$.}\label{IC-1D-PoP}
\end{figure}

We have chosen a cylindrical geometry to study the excitation of
nonlinear structures modulated by different values of the initial
azimuthal flow speeds, the distance to the symmetric axis, and the
Mach number. We not only reproduced the three familiar shapes (i.e.,
sinusoidal, sawtooth, and spiky or bipolar) of nonlinear waves
observed by satellites, but also found that nonlinear waves
supported by cylindrically symmetric plasma flows have both fast and
slow branches, both converging and diverging electric shocks, and,
both density humps and dips. This is different from the rectangular
flows. By solving Eq.(\ref{set-new2}) in a Cartesian frame, we
obtain that the wave potential $\Phi$ satisfies [c.f., e.g.,
\citet{ma09}]
\begin{equation}\begin{array}{lll}
\frac{\mathrm{d}^{2}\Phi}{\mathrm{d}X^{2}}=e^{\Phi}-\frac{\sqrt{2}}{\sqrt{(1+\frac{3\xi}{M^{2}}-\frac{2\Phi}{M^{2}})+\sqrt{\left[(1+\frac{3\xi}{M^{2}}-\frac{2\Phi}{M^{2}})\right]^{2}-12\frac{\xi}{M^{2}}}}}
\end{array}
\end{equation}
where $\xi$ is the ratio between ion and electron initial
temperatures. Taking $\xi=0.1$ and $M$=1.14,1.16,1.28, respectively,
produces a result shown in Fig.\ref{IC-1D-PoP}. Though the figure
keeps exhibiting the three familiar nonlinear structures, only
diverging shocks are seen to be driven.

\begin{figure}[htb]
\begin{center}
\includegraphics[scale=0.55,angle=270]{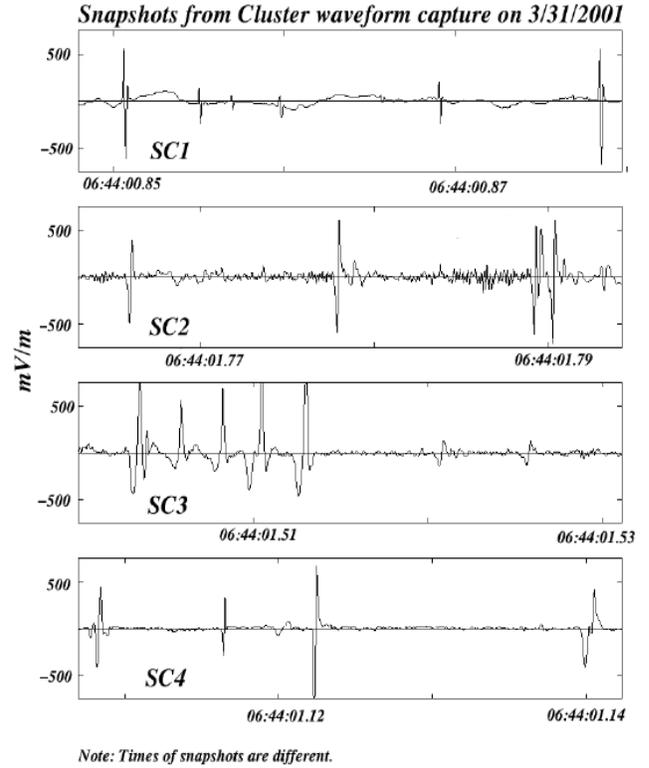}
\end{center}
\caption{Nonlinear structures detected by the four Cluster
satellites at four different locations and an $\sim$0.03 s interval
from each satellite, with amplitudes often reaching 500-750 mV/m,
the largest amplitudes measured in the outer magnetosphere. Notice
that SC1 panel illustrates converging shocks, while others record
diverging shocks. Adapted from Fig.11 of Ref.[39].}\label{Cluster}
\end{figure}

The cylindrical model and calculations presented in last Sections
are useful in gaining physical insights into the excitation and
propagation of observed nonlinear waves. Fig.\ref{Cluster} is an
example, exposing snapshots of nonlinear waves within the plasma
sheet on auroral field lines, taken by the four Cluster satellites
when they encountered the high altitude auroral zone at a radial
distance of $\sim$4.2 $R_{E}$ at 21:30MLT and $\sim62^{\circ}$ ILAT
(south) on 31 March 2001 at $\sim$06:44 UT \citep{cat03}. The panels
show both converging shocks (SC1 panel) and diverging shocks (SC2-4
panels), sawtooth structures (SC2 panel), and more complicated
shapes of background hiss, at the four locations separated by
$\sim$1000 km and an $\sim$0.03 s interval from each satellite. The
detected amplitudes of waves can reach over 750 mV/m, the largest
values ever reported in the outer magnetosphere.

At that time, the mission was experiencing a strong magnetic storm.
The magnetosphere was highly compressed due to an intense injection
of electrons $\sim$14 min earlier \citep{bak02}, which could enhance
the local magnetic field from a few mG of the background strength to
tens even hundreds of mG \citep{alf04,tju04,hal09}. Although the
Cluster has no single probe measurements to give nonlinear wave
speeds, all four panels reveal that they excellently recorded all
the bursts by double probes of high resolutions \citep{gus88,cat03}.
The pulses are $\sim$2.5 ms or its multiples in time, in good
agreement with our simulation value of $X$=10-30 by using the local
orbital data of the Cluster. More important, SC1 recorded an
oppositely propagating nonlinear wave of converging shocks,
different from other satellites which displayed diverging shocks,
but at the same time. This is in good agreement with the
illustration of two oppositely propagating nonlinear waves existing
in a cylindrical system simultaneously, as shown by
Fig.\ref{axial-3}. This fact notwithstanding, we are unable to find
the wave speed data to double-check $V_{z}$ of the wave detected by
SC1, which should have a smaller amplitude than those measured by
other satellites.

\section{Conclusion and discussion}

By using a two-fluid self-similar MHD model, we studied features of
parallel-propagating nonlinear waves driven in a cylindrically
symmetric system. We not only reproduced the three salient shapes
(i.e., sinusoidal, sawtooth, and spiky or bipolar) of
parallel-propagating, IA/IC nonlinear waves, but also found
following new results: (1) the prerequisite to trigger the nonlinear
waves is the nonzero azimuthal speed; (2) there are always two types
of nonlinear waves: a fast one propagating along $\mathbf{B}$
(diverging shocks) and a slow one against $\mathbf{B}$ (converging
shocks); (3) the distance from the symmetric axis influences the
nonlinear features: the closer to the axis, the more pronounced the
characteristics, and vice versa; (4) there exist density holes for
$M$$>$1 and humps for $M$$<$1, the boundaries of which constitute
nonlinear structures.

This study is the first step to investigate the mechanism of
nonlinear wave-particle interactions and their effects on some
unusual observed phenomena, e.g., transverse ion heating, broadband
noise emission, and magnetic holes (or bubbles, decreases
\citep{tsu05}. It offers an alternative to explain inverted-$V$
structures in beam-precipitating regions \citep{pot09}, as to be
contributed in a companion paper. This is based on the
considerations as follows.

When a field-aligned current is enclosed in the ionosphere by the
Pedersen current, plasma turbulence enhances abnormal resistance
locally which may bring about an electric field perpendicular to the
magnetic field lines. This field drives naturally an azimuthal flow,
and thus provides, on one hand, a boundary condition to
Eq.(\ref{set-new2}). On the other hand, there also exists another
magnetospheric boundary condition. Therefore, unlike the treatment
employed in the present paper, the solution that yields a
non-vanishing parallel potential drop has to be sought in a system
where Eq.(\ref{set-new2}) should be solved as a two-point boundary
condition problem, rather than an initial condition one as was the
case here. From the result of this paper, we know that the parallel
electric field decreases away from the center of the flow (see the
amplitude decrease of $E_{i}$ from Fig.\ref{axial-3} to
Fig.\ref{axial-4}). In this way, we anticipate that an alternative
solution would yield the classical inverted-$V$ structures, with
scale length comparable to the ion gyroradius.

\begin{acknowledgments}
This work is funded by Visiting Fellowships in Canadian Government
Laboratories Program, Natural Sciences and Engineering Research
Council of Canada. The author thanks Senior Scientist, W. Liu,
Canadian Space Agency, for valuable comments and suggestions.
\end{acknowledgments}

\end{document}